\documentclass[preprint,12pt]{elsarticle}
\usepackage[margin=20mm]{geometry}
\biboptions{sort&compress}
\usepackage[T1]{fontenc}
\usepackage{ragged2e}
\usepackage{amsmath,amssymb,amsfonts,mathtools,bm}
\usepackage{amsthm}
\usepackage{xfrac}
\usepackage{mathrsfs}
\usepackage{dsfont}
\usepackage{dashbox}
\usepackage{anyfontsize}
\usepackage{graphicx}
\usepackage{xcolor}
    \definecolor{notecolor}{HTML}{300d66}
    \definecolor{tmpcolor}{HTML}{AA0033}
\usepackage[colorlinks=true]{hyperref}
\usepackage{comment}
\usepackage{refcount}
\newtheorem{theorem}{Theorem}
\newtheorem{lemma}[theorem]{Lemma}

\theoremstyle{definition}

\theoremstyle{remark}

\newcommand{\appref}[1]{\hyperref[#1]{Appendix~\ref*{#1}}}
\newcommand{\secref}[1]{\hyperref[#1]{Section~\ref*{#1}}}
\newcommand{\subsecref}[1]{\hyperref[#1]{Section~\ref*{#1}}}

\newcommand{\pf}{\operatorname{pf}\mkern-1.7mu}

\newcommand{\mi}{\mathrm{i}}

\newcommand{\bdot}{\bm{\cdot}}

\newcommand{\Hir}{{\mathcal{D}}}

\newcommand{\ba}{\begin{eqnarray}}
\newcommand{\ea}{\end{eqnarray}}

\begin{document}
\hypersetup{
    linkcolor=blue,
    citecolor=blue,
    urlcolor=blue
}
\begin{frontmatter}

\title{A proof of Nakamura's conjecture}
\author[1]{A. R. Melikyan}
\ead{amelik@gmail.com}

\author[1]{A. Pinzul}
\ead{aleksandr.pinzul@gmail.com}

\affiliation[1]{
	organization={Instituto de Física, Centro Internacional de Física, \\ Universidade de Brasília},
	city={Brasília},
	state={DF},
	postcode={70910-900},
	country={Brazil}
}

\begin{abstract}
We prove Nakamura’s conjecture on the relation between the 2D Toda model and solutions of the Ernst equation, using the Pfaffian solutions obtained in our previous work \cite{Melikyan2026pf}.
\end{abstract}

\end{frontmatter}

\tableofcontents

\section{Introduction}
\label{intro}
In this paper we prove Nakamura's conjecture, concerning the relation between the 2D Toda model and Yamazaki-Hori solutions of the Ernst equation. Recall that the Ernst equation is formulated in terms of a complex potential \(\xi\), which allows reconstruction of metric functions for stationary axisymmetric gravitational fields. It has the form \cite{Ernst:1974al}:
\begin{align}
	\left( \xi \xi^{*} -1\right)\nabla^{2} \xi=2 \xi^{*}\vec{\nabla}\xi \cdot \vec{\nabla}\xi. \label{intro:Ernst_eq}
\end{align}
The Yamazaki-Hori solutions to \eqref{intro:Ernst_eq} are rational solutions parametrized by \(n \in \mathbb{Z}_{\ge 0}\), and have the form \cite{Yamazaki:1977dp,Yamazaki:1977xd}:
\begin{align}
	\xi_{n} = \frac{g_{n}(x,y,p,q)}{f_{n}(x,y,p,q)}
\end{align}
where  \(x,y\) are the prolate spheroidal coordinates and \(p,q\) are two parameters satisfying \(p^{2} +q^{2} =1 \).\footnote{In our previous work \cite{Melikyan2026pf}, the polynomials \(f_{n}\) and \(g_{n}\) were studied in the polynomial ring \(\mathbb{C}[p,q,x,y]\), with \(p\) and \(q\) treated as independent variables. In this paper, however, we impose the physical constraint \(p^{2}+q^{2}=1\) throughout.}

The semi-infinite 2D Toda molecule is described in terms of the \(\tau\)-functions \(\tau_{n}=\tau_{n}(\eta,\zeta)\) by the Hirota bilinear equation \cite{Nakamura:1993ig}:
\begin{align}
\Hir_{\eta} \Hir_{\zeta} \, \tau_{n} \bdot \tau_{n}
=
2 \tau_{n+1}\tau_{n-1}, \quad \tau_0=1. \qquad n=1,2,\ldots,
\label{intro:2d_Toda_eq_Hirota}
\end{align}
with the solutions given by the following two-directional Wronskian:
\begin{align}
    \tau_{n}(\Phi) &= \det \mathcal{T}^{(n)}(\Phi), \quad
    \mathcal{T}_{kl}^{(n)}(\Phi) := \left( \frac{\partial}{\partial \eta}
    \right)^{k-1}
              \left( \frac{\partial}{\partial \zeta}
              \right)^{l-1} \Phi(\eta,\zeta),
    \quad 1 \leq k,l \leq n.
    \label{intro:general_sol_2D_Toda}
\end{align}
Here \(\Phi(\eta,\zeta)\) is an arbitrary seed function.\footnote{The coordinates \(\eta,\zeta\) are related to the prolate spheroidal coordinates \(x,y\) by \(\eta=\frac14\ln\frac{(x-1)(1-y)}{(x+1)(1+y)}\) and \(\zeta=\frac14\ln\frac{(x-1)(1+y)}{(x+1)(1-y)}\).}

Nakamura conjectured and verified numerically for the first values of \(n\) in \cite{Nakamura:1993ig} that the polynomials \(g_{n}\) and \(f_{n}\) can be obtained from the \(\tau\)-function of the 2D Toda model, thus establishing a relation between the two models. Fukuyama, Kamimura and Yu \cite{Fukuyama:1995cm} proved the conjecture for the nonrotating Weyl case \(q=0\). Subsequently, Fukuyama and Koizumi \cite{Fukuyama:2011ma} established the conjectured relations for the highest and lowest orders in an expansion in the rotation parameter. More recently, Fukuyama \cite{Fukuyama:2026qe} investigated the underlying determinant structure and its relation to the conjecture.

In this paper we prove the conjecture for arbitrary values of the parameter \(s=q/p\),\footnote{Although the parametrization \(s=q/p\) assumes \(p\neq0\), the case \(p=0\) is recovered by continuity, since the appropriately rescaled Pfaffians are polynomials in \(p,q,x,y\), as shown in \cite{Melikyan2026pf}.} using the Pfaffian solutions to the Ernst equation obtained in \cite{Melikyan2026pf}. We recall the main result of \cite{Melikyan2026pf} in the following theorem.

\begin{theorem}[Pfaffian form of the Yamazaki-Hori solutions] \label{intro:theorem_1}
The Yamazaki-Hori solutions are expressed in terms of the polynomials \(f_{n}\) and \(g_{n}\), each given by a product of Pfaffians, while the corresponding Ernst potential \(\xi_{n}=g_{n}/f_{n}\) is given by the ratio of two Pfaffians. In particular, we have the following two cases.
\par\medskip\noindent
{Even case (\(m=n-1=2k\)):}
\begin{align}
    f_{n} &= C^{(f)} \, \overline{\pf \big(\omega^{(s)}\big)} \pf \big(\omega^{(s)}\big),
    \quad C^{(f)}= \frac{(-1)^{n} p^{2n-2} C(n)}{2n-1}, \\
    g_{n} &= C^{(g)}\, \overline{\pf \big(\omega^{(s)}\big)} \pf \big(\Omega^{(e)}\big),
 \quad C^{(g)}= (-1)^{n} p^{2n-1} C(n).
\end{align}
Thus, for the Ernst potential:
\begin{equation}
    \xi_{n} = p(2n-1) \frac{\pf \big(\Omega^{(e)}\big)} {\pf \big(\omega^{(s)}\big)}.
\end{equation}
\par\medskip\noindent
{Odd case (\(m=n-1=2k+1\)):}
\begin{align}
    f_{n} &= \widetilde{C}^{(f)}\, \overline{\pf \big(\widehat{\omega}^{(s)}\big)} \pf \big(\widehat{\omega}^{(s)}\big), \quad \widetilde{C}^{(f)} = \frac{(-1)^{n} p^{2n-2} C(n)}{2(2n-1)},\\
    g_{n} &= \widetilde{C}^{(g)}\, \overline{\pf \big(\widehat{\omega}^{(s)}\big)} \pf \big(\Omega^{(o)}\big), \quad \widetilde{C}^{(g)} =(-1)^{n} p^{2n-1}C(n).
\end{align}
Thus, for the Ernst potential:
\begin{equation}
    \xi_{n} = 2p(2n-1)
    \frac{\pf \big(\Omega^{(o)}\big)}
         {\pf \big(\widehat{\omega}^{(s)}\big)}.
\end{equation}
In these formulas, the constant \(C(n)\) is given by the following expression:
\begin{align}
	C(n):=\prod_{j=1}^{n-1}\big[-2(2j-1)\big] =(-1)^{n-1}\frac{\Gamma(2n-1)}{\Gamma(n)}, \label{intro:C_n_def}
\end{align}
and the explicit forms of the Pfaffians in the above formulas are collected in \ref{app:appa}.
\end{theorem}

We use the above theorem in order to prove Nakamura's conjecture, which we formulate as follows:
\begin{theorem}[Nakamura's conjecture] \label{intro:theorem_2}
The Yamazaki-Hori Ernst potential is related to the \(\tau\)-functions of the 2D Toda model by:
\begin{equation}
\xi_{n}=-\frac{\tau_{n}(\Phi)}{\tau_{n-1}(\Phi')},\qquad n\geq1.
\end{equation}
More precisely, setting \(m=n-1\), the following identities hold:
\par\medskip\noindent
Even case (\(m=2k\)):
\begin{align}
    \tau_{m}(\Phi') &= \mathcal{N}_{m}\,\pf\big(\omega^{(s)}\big),\\
    \tau_{n}(\Phi) &= -p(2m+1)\mathcal{N}_{m}\,\pf\big(\Omega^{(e)}\big).
\end{align}
\par\medskip\noindent
Odd case (\(m=2k+1\)):
\begin{align}
    \tau_{m}(\Phi') &= \mathcal{N}_{m}\,\pf\big(\widehat{\omega}^{(s)}\big),\\
    \tau_{n}(\Phi) &= -2p(2m+1)\mathcal{N}_{m}\,\pf\big(\Omega^{(o)}\big).
\end{align}
Here the normalization factor \(\mathcal{N}_{m}\) is given by
\begin{equation}
\mathcal{N}_{m}=2^{\lfloor m/2\rfloor}p^{m}\prod_{j=1}^{m}(2j-1)(2j)!,\qquad \mathcal{N}_{0}=1,
\end{equation}
where \(\lfloor m/2\rfloor\) denotes the integer part of \(m/2\). The two seed functions \(\Phi\) and \(\Phi'=\partial_{\zeta}\partial_{\eta}\Phi\) are given by:
\begin{align}
\Phi(x,y)&=-px+\mi qy,\label{intro:seed_Phi}\\
\Phi'(x,y)&=-2px(x^{2}-1)-2\mi qy(y^{2}-1).\label{intro:seed_Phi_prime}
\end{align}
\end{theorem}

\section{Proof of Nakamura's conjecture}
\label{sec:nc}

\subsection{Polynomial identities and the Toda matrix}
First, we consider the even case corresponding to \(m=n-1=2k\), and the polynomial \(f_{n}\). According to Theorem \ref{intro:theorem_1}:
\begin{align}
f_{n} &= C^{(f)} \, \overline{\pf \big(\omega^{(s)}\big)} \pf \big(\omega^{(s)}\big),
\quad C^{(f)}= \frac{(-1)^n p^{2n-2} C(n)}{2n-1},
\end{align}
where (see \ref{app:appa} for details):
\begin{align}
    &\omega^{(s)}_{ij}= \omega_{ij} - \frac{\mi s}{2}  \left[\Gamma_{i}(x)\Gamma_{j}(y) -\Gamma_{i}(y)\Gamma_{j}(x) \right], \quad \Gamma_{k}(w):= \frac{1}{2k-1}\left[w^{2k-1}-w^{2k+1}\right] \label{intro:omegas_Gamma}\\
	&\omega_{ij}= \frac{1}{2} \mathcal{I}_{ij}(x) +\frac{s^{2}}{2}\mathcal{I}_{ij}(y), \quad \mathcal{I}_{ij}(z):=\int_{1}^{z} \left[\Gamma_{i}(w) \Gamma_{j}^{\prime}(w) -\Gamma_{i}^{\prime}(w) \Gamma_{j}(w) \right] \, dw
\end{align}
We now relate the polynomials \(\Gamma_{k}(z)\) to the polynomials \(P_{n}(z)\), which arise in the Toda \(\tau\)-function associated with the seed \(\Phi'\) and are defined by the following recurrence relation:\footnote{The polynomials \(P_{n}\) are closely related to the Eulerian polynomials \(A_{n}\). Indeed, setting \(z=(1+t^{2})/(1-t^{2})\), we obtain \(P_{n}(z)=2^{n+1}\psi_{n}(t^{2})\), for \(n\geq1\), where \(\psi_{n}(w)=\sum_{r=1}^{\infty}r^{n}w^{r}=A_{n}(w)/(1-w)^{n+1}\) (see \cite{Vein:2013kv} for details).}

\begin{align}
P_{k+1}(z)=(z^2-1)\frac{d}{dz}P_{k}(z), \quad P_{0}(z)=z, \label{intro:Pn_def}
\end{align}
For the seed function \(\Phi'\) \eqref{intro:seed_Phi_prime}, one finds:\footnote{The relation between the derivatives in \((\eta,\zeta)\) and \((x,y)\) coordinates is given by:
\begin{align*}
\partial_{\eta} &=(x^{2}-1)\partial_{x}+(y^{2}-1)\partial_{y},\\
\partial_{\zeta} &=(x^{2}-1)\partial_{x}-(y^{2}-1)\partial_y.
\end{align*}
}
\begin{align}
{\partial_\eta}^{a} {\partial_\zeta}^{b}\Phi' = -p\left[P_{a+b+2}(x)+\mi s (-1)^{b}P_{a+b+2}(y)\right].
\end{align}
Denoting:
\begin{align}
U_{j} &= P_{j+2}(x)+\mi s P_{j+2}(y),\\
V_{j} &= P_{j+2}(x)-\mi s P_{j+2}(y).
\end{align}
the \(\tau\)-function of the 2D Toda model can be written in the form:
\begin{equation}
\tau_{2k}(\Phi')=(-p)^{2k}\det\mathcal{T}_{2k}^{(\Phi')},
\end{equation}
where the \(2k\times 2k\) Toda matrix \(\mathcal{T}_{2k}^{(\Phi')}\) has the form:

\begin{equation}
\mathcal{T}_{2k}^{(\Phi')}=
\begin{pmatrix}
U_{0} & V_{1} & U_{2} & V_{3} & \cdots & U_{m-2} & V_{m-1}\\
U_{1} & V_{2} & U_{3} & V_{4} & \cdots & U_{m-1} & V_{m}\\
U_{2} & V_{3} & U_{4} & V_{5} & \cdots & U_{m} & V_{m+1}\\
\vdots & \vdots & \vdots & \vdots & \ddots & \vdots & \vdots\\
U_{m-2} & V_{m-1} & U_{m} & V_{m+1} & \cdots & U_{2m-4} & V_{2m-3}\\
U_{m-1} & V_{m} & U_{m+1} & V_{m+2} & \cdots & U_{2m-3} & V_{2m-2}
\end{pmatrix},\qquad m=2k.
\label{nc:Toda_matrix}
\end{equation}
Noting that \(\deg P_j=j+1\) and \(\deg\Gamma_j=2j+1\), and writing \(P_{2j-1}(z)=\sum_{\ell=0}^{j}a_{2\ell}^{(j)}z^{2\ell}\), for \(j=1,\ldots,m\), it follows directly from the recurrence relation \eqref{intro:Pn_def} that the two polynomial sequences are related by:
\begin{align}
    &P_{2j}(z)=\sum_{\ell=1}^{j}\Lambda_{j\ell}^{(m)}\Gamma_\ell(z), \qquad j=1,\ldots,m, \label{nc:P_Gamma_rel}
    \\
    &\Lambda_{j\ell}^{(m)}=-2\ell(2\ell-1)a_{2\ell}^{(j)}, \qquad 1\leq\ell\leq j\leq m. \label{nc:Lambda_def}
\end{align}
Thus, \(\Lambda^{(m)}\) is a lower-triangular matrix. Since \(a_{2j}^{(j)}=(2j-1)!\), we find:
\begin{align}
	\det\Lambda^{(m)}=(-1)^{m}\prod_{\ell=1}^{m}(2\ell)!(2\ell-1)
\end{align}

With the explicit formulas relating the polynomial sequences \(P_{n}\) and \(\Gamma_{n}\), we can evaluate \(\omega^{(s)}\) in terms of the polynomials \(P_{n}\), which is the first step to compare \(\pf \big(\omega^{(s)} \big)\) with the determinant of the matrix \(\mathcal{T}_{2k}^{(\Phi')}\). Defining (cf \eqref{intro:omegas_Gamma}):
\begin{align}
\mathcal{I}_{ij}^{(P)}(z):=\int_{1}^{z} \left[P_{i}(w) P_{j}^{\prime}(w) -P_{i}^{\prime}(w) P_{j}(w) \right] \, dw
\end{align}
and using the relation between \(P_{2i}\) and \(\Gamma_{j}\) \eqref{nc:P_Gamma_rel}, we find:
\begin{align}
\mathcal{I}^{(P)}_{2r,2s}(z)=\sum_{i,j=1}^{m}\Lambda^{(m)}_{ri}\Lambda^{(m)}_{sj}\mathcal{I}_{ij}(z), \qquad 1 \leq r,s \leq m.
\end{align}
Integration by parts and the recurrence relation \eqref{intro:Pn_def} give:
\begin{align}\label{IPP}
	\mathcal{I}^{(P)}_{rs}(z)=P_{r}(z)P_{s}(z) - P_{r+1}(z)P_{s-1}(z) -\mathcal{I}^{(P)}_{r+1, s-1}(z).
\end{align}
Applying repeatedly the above formula, and noting that for even \(s-r\), the iteration terminates with \(\mathcal{I}^{(P)}_{(r+s)/2, (r+s)/2}(z)=0\), one obtains (note that the products \(P_{r+j}P_{s-j}\) with \(1\leq j<(s-r)/2\) occur twice, while \(P_{r}P_{s}\) and \(P_{(r+s)/2}^{2}\) occur only once):
\begin{align}
\mathcal{I}^{(P)}_{rs}(z)=\sum_{j=0}^{s-r-1}(-1)^{j}P_{r+j}(z)P_{s-j}(z),\qquad 0\leq r<s,\quad s-r\ \text{even}.
\label{nc:I_P_alt}
\end{align}
Using the above formula, and the definitions of \(U_j\) and \(V_j\), we find:
\begin{align}
2\sum_{i,j=1}^{m}\Lambda^{(m)}_{ai}\Lambda^{(m)}_{bj}\omega^{(s)}_{ij}
=\sum_{\ell=0}^{2(b-a)-1}(-1)^{\ell} U_{2a-2+\ell}V_{2b-2-\ell},
\qquad 1\leq a<b\leq m.\label{nc:pre_Cusick}
\end{align}

\subsection{Extended Cusick identities}

The following generalized Cusick identity was established in \cite{Vein:2013kv}:
\begin{theorem}[Generalized Cusick identity]
\label{nc:Dale_Cusick_theorem}
Let \(\phi_{r}, \psi_{r}, \, r \geq 1\) be arbitrary sequences, and \(\mathcal{C}(\phi,\psi)=(c_{ab})\) the antisymmetric matrix of order \(2k\), with entries:
\begin{align}
c_{ab}=\sum_{\ell=1}^{b-a}\phi_{a+\ell-1}\psi_{b-\ell}, \qquad 1\leq a<b\leq 2k.
\end{align}
The generalized Cusick identity states that:
\begin{align}
\pf \mathcal{C}(\phi,\psi)=\det\big(\phi_{i+j+1}\big)_{i,j=0}^{k-1}\det\big(\psi_{i+j+1}\big)_{i,j=0}^{k-1}.
\end{align}
\end{theorem}
\noindent Setting \(u_j=\phi_{j+1}\), \(v_j=\psi_{j+1}\), and shifting the summation index \(\ell\to\ell+1\), we define the antisymmetric matrix \(\mathcal{C}(u,v)\) of arbitrary order \(m\) by
\begin{equation}
\mathcal{C}(u,v)_{ab}:=\sum_{\ell=0}^{b-a-1}u_{a-1+\ell} \, v_{b-2-\ell},\qquad 1\leq a<b\leq m,
\end{equation}
with antisymmetric extension to \(a>b\). In this notation, the generalized Cusick identity becomes:
\begin{align}
\pf\mathcal{C}(u,v)=\det\mathcal{U}\,\det\mathcal{V},
\end{align}
where \(\mathcal{U}_{ij}=u_{i+j}\) and \(\mathcal{V}_{ij}=v_{i+j}\), for \(0\leq i,j<k\).

We now show that the sum in \eqref{nc:pre_Cusick} can be expressed as a difference of two generalized Cusick sums. To this end, we introduce four sequences:
\begin{align}
u_j=U_{2j},\qquad p_j=V_{2j+1},\qquad q_j=U_{2j+1},\qquad v_j=V_{2j+2}.
\end{align}
Then, the equation \eqref{nc:pre_Cusick} becomes:
\begin{align}
2\sum_{i,j=1}^{m}\Lambda^{(m)}_{ai}\Lambda^{(m)}_{bj}\omega^{(s)}_{ij}
=\sum_{\ell=0}^{b-a-1}\left(u_{a-1+\ell}v_{b-2-\ell}-q_{a-1+\ell}p_{b-2-\ell}\right), \qquad 1\leq a<b\leq m.
\label{nc:cusick_paired}
\end{align}
Using the definition of \(\mathcal{C}(u,v)\), we write the formula \eqref{nc:cusick_paired} in the matrix form:\footnote{
Since the derivation does not depend on the parity of \(m\), the relation \eqref{nc:omega_cusick_relation} holds for both even and odd \(m\).
}
\begin{align}
2\Lambda^{(m)}\omega^{(s)}\big(\Lambda^{(m)}\big)^T=\mathcal{C}(u,v)-\mathcal{C}(q,p).
\label{nc:omega_cusick_relation}
\end{align}
This relation motivates the following extension of Dale's generalized Cusick identity to four arbitrary sequences.
\begin{theorem}[Extended Cusick identity]
\label{naka:ext_Cusick}
Let \(m=2k\), let \(u_{j},p_{j},q_{j},v_{j}\), \(0 \leq j \leq m-2\), be four arbitrary sequences, and let \(\mathcal{C}(u,v)\) and \(\mathcal{C}(q,p)\) be antisymmetric matrices of order \(m\), defined as above. Then:
\begin{align}
\pf\left[\mathcal{C}(u,v)-\mathcal{C}(q,p)\right]
=\det\begin{pmatrix}
\mathcal{U}&\mathcal{P}\\
\mathcal{Q}&\mathcal{V}
\end{pmatrix},
\label{nc:extended_Cusick}
\end{align}
where \(\mathcal{U},\mathcal{P},\mathcal{Q},\mathcal{V}\) are \(k\times k\) Hankel matrices with entries
\begin{align}
\mathcal{U}_{ij}=u_{i+j},\qquad \mathcal{P}_{ij}=p_{i+j},\qquad \mathcal{Q}_{ij}=q_{i+j},\qquad \mathcal{V}_{ij}=v_{i+j}, \qquad 0\leq i,j<k.
\end{align}
\end{theorem}
\noindent Setting \(p_{j}=q_{j}=0\), the determinant on the right-hand side becomes block diagonal, and we recover the generalized Cusick identity of \cite{Vein:2013kv}. The extended Cusick identity also implies a corresponding bordered identity for odd \(m=2k+1\).
\begin{theorem}[Bordered extended Cusick identity]
\label{naka:bordered_Cusick}
Let \(m=2k+1\), let \(u_{j}\), \(0\leq j\leq 2k\), and \(p_{j},q_{j},v_{j}\), \(0\leq j\leq 2k-1\), be arbitrary sequences, and let \(C_m=\mathcal{C}(u,v)-\mathcal{C}(q,p)\) be the antisymmetric matrix of order \(m\), defined as above. Introducing the column vector
\begin{equation}
\mathbf{u}=(u_{0},u_{1},\ldots,u_{m-1})^{T},
\end{equation}
we have
\begin{equation}
\pf\begin{pmatrix}C_{m} & \mathbf{u} \\
-\mathbf{u}^{T} & 0\end{pmatrix}=\det\begin{pmatrix}\mathcal{U} & \mathcal{P}\\
\mathcal{Q} & \mathcal{V}\end{pmatrix}, \label{nc:bordered_cusick}
\end{equation}
where the four Hankel blocks \(\mathcal{U},\mathcal{P},\mathcal{Q},\mathcal{V}\) have dimensions \((k+1)\times(k+1)\), \((k+1)\times k\), \(k\times(k+1)\), and \(k\times k\), respectively, with entries:
\begin{alignat}{2}
\mathcal{U} &= \big(u_{i+j}\big)_{i,j=0}^{k}, &\quad \mathcal{P} &= \big(p_{i+j}\big)_{\substack{0\le i\le k\\0\le j<k}},\\
\mathcal{Q} &= \big(q_{i+j}\big)_{\substack{0\le i<k\\0\le j\le k}}, &\quad \mathcal{V} &= \big(v_{i+j}\big)_{i,j=0}^{k-1}.
\end{alignat}
\end{theorem}
\noindent The proofs of both extended Cusick identities are given in \ref{app:appb}.

Nakamura's conjecture follows from the two extended Cusick identities stated above.
We consider each case corresponding to the seed functions \(\Phi'\) and \(\Phi\) separately.

\subsection{Proof for the seed \(\Phi'\) }
\begin{lemma}
Let \(m=n-1\), and consider the Toda \(\tau\)-function associated with the seed \(\Phi'\). Then:\footnote{Recall that \(\tau_{\ell}(\Psi)=(-p)^{\ell}\det\mathcal{T}_{\ell}^{(\Psi)}\), for any \(\ell\) and \(\Psi=\Phi,\Phi'\).}
\begin{equation}
\det\mathcal{T}_{m}^{(\Phi')}=
\begin{cases}
2^{k}\det\Lambda^{(m)}\pf \big(\omega^{(s)}\big), & m=2k,\\
2^{k}\det\Lambda^{(m)}\pf \big(\widehat\omega^{(s)}\big), & m=2k+1
\end{cases},
\end{equation}
where \(\Lambda^{(m)}\) is the lower-triangular matrix defined in \eqref{nc:Lambda_def}, and \(\omega^{(s)}, \, \widehat\omega^{(s)}\) are antisymmetric matrices defined in \ref{app:appa}.
\end{lemma}
\begin{proof}
\leavevmode\par
\noindent\textbf{Even case (\(m=2k\)).} By elementary row and column permutations, the Toda matrix \(\mathcal{T}_{2k}^{(\Phi')}\) \eqref{nc:Toda_matrix} can be brought into the block-Hankel form:
   \begin{align}
   \det\mathcal{T}_{2k}^{(\Phi')}
   =\det\begin{pmatrix}
   \mathcal{U} & \mathcal{P}\\
   \mathcal{Q} & \mathcal{V}
   \end{pmatrix},
   \end{align}
   where, for \(0\le i,j<k\),
   \begin{equation}
   \begin{aligned}
   \mathcal{U}_{ij}&=U_{2(i+j)}, & \mathcal{P}_{ij}&=V_{2(i+j)+1},\\
   \mathcal{Q}_{ij}&=U_{2(i+j)+1}, & \mathcal{V}_{ij}&=V_{2(i+j)+2}.
   \end{aligned}
   \end{equation}
   Thus, the four sequences in the extended Cusick identity are:
   \begin{equation}
   u_j=U_{2j},\qquad p_j=V_{2j+1},\qquad q_j=U_{2j+1},\qquad v_j=V_{2j+2}.
   \end{equation}
   Then, applying the extended Cusick identity and using \eqref{nc:omega_cusick_relation}, we find
   \begin{align}
   \det\mathcal{T}_{2k}^{(\Phi')}=\pf\left[\mathcal{C}(u,v)-\mathcal{C}(q,p)\right]
   =\pf\left[2\Lambda^{(m)}\omega^{(s)}\big(\Lambda^{(m)}\big)^T\right]
   =2^k\det \big(\Lambda^{(m)}\big)\,\pf \big(\omega^{(s)} \big).
   \end{align}

  \par\medskip\noindent\textbf{Odd case (\(m=2k+1\)).} The corresponding Toda matrix has the form:
   \begin{equation}
   \mathcal{T}_{2k+1}^{(\Phi')}=
   \left(
   \begin{array}{ccccc}
   U_{0} & V_{1} & U_{2} & \cdots & U_{m-1} \\
   U_{1} & V_{2} & U_{3} & \cdots & U_{m} \\
   \vdots & \vdots & \vdots & \ddots & \vdots \\
   U_{m-1} & V_{m} & U_{m+1} & \cdots & U_{2m-2}
   \end{array}
   \right),
   \qquad m=2k+1.
   \end{equation}
   The definitions of \(u,p,q,v\) and the relation \eqref{nc:omega_cusick_relation} are valid for arbitrary \(m\), so we now use them with \(m=2k+1\). Then as in the even case, by elementary row and column permutations, the Toda matrix \(\mathcal{T}_{m}^{(\Phi')}\) can be brought into the block-Hankel form. Applying the bordered extended Cusick identity, we obtain
   \begin{equation}
   \det\mathcal{T}_{m}^{(\Phi')}=\det\begin{pmatrix}\mathcal{U} & \mathcal{P}\\\mathcal{Q} & \mathcal{V}\end{pmatrix}=\pf\begin{pmatrix}\mathcal{C}(u,v)-\mathcal{C}(q,p) & \mathbf{u}\\-\mathbf{u}^{T} & 0\end{pmatrix},
   \end{equation}
   where \(\mathbf{u}=\Lambda^{(m)}\boldsymbol\Gamma^{(+)}=(u_{0},\ldots,u_{m-1})^{T}\). Using \eqref{nc:omega_cusick_relation}, we find:
   \begin{equation}
   \det\mathcal{T}_{m}^{(\Phi')}=\pf\begin{pmatrix}2\Lambda^{(m)}\omega^{(s)}\big(\Lambda^{(m)}\big)^{T} & \Lambda^{(m)}\boldsymbol\Gamma^{(+)}\\-\big(\boldsymbol\Gamma^{(+)}\big)^{T}\big(\Lambda^{(m)}\big)^{T} & 0\end{pmatrix}.
   \end{equation}
   Defining \(D=\operatorname{diag}(\Lambda^{(m)},1)\), we obtain:
   \begin{equation}
   \det\mathcal{T}_{m}^{(\Phi')}=\pf\left[D\begin{pmatrix}2\omega^{(s)}&\boldsymbol{\Gamma}^{(+)}\\-(\boldsymbol{\Gamma}^{(+)})^{T}&0\end{pmatrix}D^{T}\right]
   =2^{k}\det\Lambda^{(m)}\,\pf\begin{pmatrix}\omega^{(s)}&\boldsymbol{\Gamma}^{(+)}\\-(\boldsymbol{\Gamma}^{(+)})^{T}&0\end{pmatrix}.
   \end{equation}
   Using the relation between \(\omega\) and \(\omega^{(s)}\):
   \begin{equation}
   \omega=\omega^{(s)}+\frac{1}{2}\left[\boldsymbol\Gamma(x)(\boldsymbol\Gamma^{(+)})^{T}-\boldsymbol\Gamma^{(+)}\boldsymbol\Gamma(x)^{T}\right],
   \end{equation}
   and introducing the triangular matrix:
   \begin{equation}
   L=\begin{pmatrix}I_{m} & -\frac{1}{2}\boldsymbol\Gamma(x)\\0 & 1\end{pmatrix},\qquad \det L=1,
   \end{equation}
   we obtain:
   \begin{equation}
   \widehat\omega^{(s)}=L\begin{pmatrix}\omega^{(s)} & \boldsymbol\Gamma^{(+)}\\-(\boldsymbol\Gamma^{(+)})^{T} & 0\end{pmatrix}L^{T}.
   \end{equation}
   Since \(\det L=1\), the Pfaffian identity \(\pf(MAM^{T})=\det(M)\pf(A)\) gives:
   \begin{equation}
   \pf\big(\widehat\omega^{(s)}\big)=\pf\begin{pmatrix}\omega^{(s)} & \boldsymbol\Gamma^{(+)}\\-(\boldsymbol\Gamma^{(+)})^{T} & 0\end{pmatrix}.
   \end{equation}
   Therefore,
   \begin{equation}
  \det\mathcal{T}_{m}^{(\Phi')}=2^{k}\det\Lambda^{(m)}\,\pf\big(\widehat\omega^{(s)}\big),\qquad m=2k+1.
   \end{equation}
\end{proof}

Next, we prove analogous results for the seed \(\Phi\).

\subsection{Proof for the seed \(\Phi\)}
\begin{lemma}
Let \(m=n-1\), and consider the Toda \(\tau\)-function associated with the seed \(\Phi\). Then
\begin{equation}
\det\mathcal{T}_{m+1}^{(\Phi)}=
\begin{cases}
2^{k}(2m+1)\det\Lambda^{(m)}\pf\big(\Omega^{(e)}\big), & m=2k,\\
2^{k+1}(2m+1)\det\Lambda^{(m)}\pf\big(\Omega^{(o)}\big), & m=2k+1,
\end{cases}
\end{equation}
where \(\Lambda^{(m)}\) is defined in \eqref{nc:Lambda_def}, and \(\Omega^{(e)},\Omega^{(o)}\) are given in \ref{app:appa}.
\end{lemma}
\begin{proof}
\leavevmode\par
\noindent\textbf{Even case (\(m=2k\)).} For the seed \(\Phi=-px+\mi qy\), we introduce the sequences
\begin{equation}
A_{j}=P_{j}(x)-\mi sP_{j}(y),\qquad B_{j}=P_{j}(x)+\mi sP_{j}(y).
\end{equation}
The corresponding normalized Toda matrix has the form:
\begin{equation}
\mathcal{T}_{m+1}^{(\Phi)}=
\begin{pmatrix}
A_0&B_1&A_2&\cdots&A_m\\
A_1&B_2&A_3&\cdots&A_{m+1}\\
\vdots&\vdots&\vdots&\ddots&\vdots\\
A_m&B_{m+1}&A_{m+2}&\cdots&A_{2m}
\end{pmatrix},\qquad m=2k.
\end{equation}
As in the denominator case, by elementary row and column permutations, the Toda matrix can be brought into block-Hankel form:
\begin{equation}
\det\mathcal{T}_{m+1}^{(\Phi)}=\det\begin{pmatrix}\mathcal{U}&\mathcal{P}\\\mathcal{Q}&\mathcal{V}\end{pmatrix},
\end{equation}
where
\begin{equation}
\begin{aligned}
\mathcal{U}_{ij}&=A_{2(i+j)}, & \mathcal{P}_{ij}&=B_{2(i+j)+1},\\
\mathcal{Q}_{ij}&=A_{2(i+j)+1}, & \mathcal{V}_{ij}&=B_{2(i+j)+2},
\end{aligned}
\end{equation}
with block dimensions \((k+1)\times(k+1)\), \((k+1)\times k\), \(k\times(k+1)\), and \(k\times k\), respectively. Thus, the four sequences entering the extended Cusick identity are
\begin{equation}
u_j=A_{2j},\qquad p_j=B_{2j+1},\qquad q_j=A_{2j+1},\qquad v_j=B_{2j+2}.
\end{equation}
Applying the bordered extended Cusick identity, we obtain
\begin{equation}
\det\mathcal{T}_{m+1}^{(\Phi)}=\pf\begin{pmatrix}\mathcal{C}(u,v)-\mathcal{C}(q,p)&\mathbf{u}\\-\mathbf{u}^{T}&0\end{pmatrix},
\end{equation}
where \(\mathbf{u}=(A_0,A_2,\ldots,A_{2m})^{T}\).

To express this Pfaffian in terms of the Ernst matrices, we first note that the same calculation leading to \eqref{nc:pre_Cusick} gives:\footnote{Indeed, let \(r=2a-2\), \(t=2b-2\), and \(d=t-r=2(b-a)\). Substituting the definitions of \(A_{j}\) and \(B_{j}\), the terms involving only \(x\) or \(y\) give \(\mathcal{I}_{r,t}^{(P)}(x)+s^{2}\mathcal{I}_{r,t}^{(P)}(y)\). For the mixed terms, reversing the summation index \(\ell\to d-\ell\) gives
\begin{equation*}
\sum_{\ell=0}^{d-1}(-1)^{\ell}P_{r+\ell}(y)P_{t-\ell}(x)=\sum_{\ell=1}^{d}(-1)^{\ell}P_{r+\ell}(x)P_{t-\ell}(y).
\end{equation*}
Thus, since \(d\) is even, all intermediate terms cancel, leaving \(P_{r}(x)P_{t}(y)-P_{r}(y)P_{t}(x)\). We have also used the fact that the identity \eqref{nc:I_P_alt} remains valid for \(r=0\), since \(P_{j}(1)=0\) for \(j\geq1\), and therefore the boundary terms at \(z=1\) vanish.}
\begin{align}
&\big[\mathcal{C}(u,v) -\mathcal{C}(q,p)\big]_{ab}
=\sum_{\ell=0}^{2(b-a)-1}(-1)^{\ell}A_{2a-2+\ell}B_{2b-2-\ell}\nonumber\\
&=\mathcal{I}^{(P)}_{2a-2,2b-2}(x)+s^{2}\mathcal{I}^{(P)}_{2a-2,2b-2}(y)+\mi s\left[P_{2a-2}(x)P_{2b-2}(y)-P_{2a-2}(y)P_{2b-2}(x)\right],\quad 1\leq a<b\leq m+1.
\label{nc:Phi_Cusick_sum}
\end{align}
Note that unlike the denominator case, the polynomial sequence now begins with \(P_0(z)=z\). From the definition of \(\Gamma_j\), we have:
\begin{equation}
P_0(z)=\sum_{j=1}^{m}(2j-1)\Gamma_j(z)+(2m+1)F(z),\qquad F(z)=\frac{z^{2m+1}}{2m+1}.
\end{equation}
Introducing the vector
\begin{equation}
\mathbf{d}=(1,3,\ldots,2m-1)^{T},
\end{equation}
and using \eqref{nc:P_Gamma_rel}, we obtain
\begin{equation}
\begin{pmatrix}P_{0}\\P_{2}\\\vdots\\P_{2m}\end{pmatrix}
=M^{(m)}\begin{pmatrix}\Gamma_{1}\\\vdots\\\Gamma_{m}\\F\end{pmatrix},
\qquad
M^{(m)}=\begin{pmatrix}\mathbf{d}^{T}&2m+1\\\Lambda^{(m)}&\mathbf{0}\end{pmatrix}.\label{nc:polynom_transform_numerator}
\end{equation}
Expanding along the last column, we find
\begin{equation}
\det M^{(m)}=(-1)^m(2m+1)\det\Lambda^{(m)}.
\label{nc:Phi_M_det}
\end{equation}
Substituting the polynomial transformation \eqref{nc:polynom_transform_numerator} into \eqref{nc:Phi_Cusick_sum}, the terms involving \(\Gamma_{i}\) and \(\Gamma_{j}\) give:
\begin{align}
&\int_{1}^{x}(\Gamma_{i}\Gamma_{j}'-\Gamma_{i}'\Gamma_{j})\,dw+s^{2}\int_{1}^{y}(\Gamma_{i}\Gamma_{j}'-\Gamma_{i}'\Gamma_{j})\,dw+\mi s[\Gamma_{i}(x)\Gamma_{j}(y)-\Gamma_{i}(y)\Gamma_{j}(x)]\nonumber\\
&=2\omega^{(-s)}_{ij}.
\end{align}
Similarly, the terms involving \(\Gamma_{i}\) and \(F\), after integration by parts using \(F'(w)=w^{2m}\) and \(\Gamma_{i}(1)=0\), give:
\begin{align}
&\int_{1}^{x}(\Gamma_{i}F'-\Gamma_{i}'F)\,dw+s^{2}\int_{1}^{y}(\Gamma_{i}F'-\Gamma_{i}'F)\,dw+\mi s[\Gamma_{i}(x)F(y)-\Gamma_{i}(y)F(x)]\nonumber\\
&=2\left[\int_{1}^{x}\Gamma_{i}(w)w^{2m}\,dw+s^{2}\int_{1}^{y}\Gamma_{i}(w)w^{2m}\,dw\right]-\gamma\overline{\beta_{i}}\nonumber\\
&=2\rho_{i}-\gamma\overline{\beta_{i}}=-2\sigma_{i},
\end{align}
where the quantities \(\rho_{i}\), \(\beta_{i}\), \(\gamma\), and \(\sigma_{i}\) are defined in \ref{app:appa}. Thus, \eqref{nc:Phi_Cusick_sum} takes the matrix form:
\begin{equation}
\mathcal{C}(u,v)-\mathcal{C}(q,p)
=2M^{(m)}\begin{pmatrix}\omega^{(-s)}&-\boldsymbol{\sigma}\\\boldsymbol{\sigma}^{T}&0\end{pmatrix}\big(M^{(m)}\big)^{T},
\label{nc:Phi_Cusick_relation}
\end{equation}
where \(\boldsymbol{\sigma}=(\sigma_1,\ldots,\sigma_m)^{T}\). The polynomial transformation \eqref{nc:polynom_transform_numerator} also gives the border vector:
\begin{equation}
\mathbf{u}=M^{(m)}\begin{pmatrix}\boldsymbol{\beta}\\\gamma\end{pmatrix},\qquad \boldsymbol{\beta}=(\beta_1,\ldots,\beta_m)^{T}.
\end{equation}
Introducing \(D=\operatorname{diag}(M^{(m)},1)\), we then find:
\begin{align}
\det\mathcal{T}_{m+1}^{(\Phi)}=\pf\left[D\begin{pmatrix}2\omega^{(-s)}&-2\boldsymbol{\sigma}&\boldsymbol{\beta}\\2\boldsymbol{\sigma}^{T}&0&\gamma\\-\boldsymbol{\beta}^{T}&-\gamma&0\end{pmatrix}D^{T}\right]=2^k\det M^{(m)}\,\pf\begin{pmatrix}\omega^{(-s)}&-\boldsymbol{\sigma}&\boldsymbol{\beta}\\\boldsymbol{\sigma}^{T}&0&\gamma\\-\boldsymbol{\beta}^{T}&-\gamma&0\end{pmatrix}.
\end{align}
Now, from the definitions of \(\omega^{(s)}\) and \(\sigma_i\), we have
\begin{equation}
\omega^{(s)}=\omega^{(-s)}+\frac12\left[\overline{\boldsymbol{\beta}}\boldsymbol{\beta}^{T}-\boldsymbol{\beta}\overline{\boldsymbol{\beta}}^{T}\right],\qquad -\boldsymbol{\sigma}=\boldsymbol{\rho}-\frac{\gamma}{2}\overline{\boldsymbol{\beta}}.
\end{equation}
Introducing the triangular matrix
\begin{equation}
L=\begin{pmatrix}I_m&0&-\frac12\overline{\boldsymbol{\beta}}\\0&1&0\\0&0&1\end{pmatrix},\qquad \det L=1,
\end{equation}
we obtain
\begin{equation}
\Omega^{(e)}=L\begin{pmatrix}\omega^{(-s)}&-\boldsymbol{\sigma}&\boldsymbol{\beta}\\\boldsymbol{\sigma}^{T}&0&\gamma\\-\boldsymbol{\beta}^{T}&-\gamma&0\end{pmatrix}L^{T}.
\end{equation}
Since \(\det L=1\), the Pfaffian is unchanged by this transformation. Using \eqref{nc:Phi_M_det}, we conclude that
\begin{equation}
\det\mathcal{T}_{m+1}^{(\Phi)}=2^k(2m+1)\det\Lambda^{(m)}\,\pf\big(\Omega^{(e)}\big),\qquad m=2k.
\end{equation}

\par\medskip
\noindent\textbf{Odd case (\(m=2k+1\)).} The Toda matrix now has even order \(m+1=2k+2\). Using the same definitions of \(u,p,q,v\) as in the even case, we can bring it into block-Hankel form, with all four blocks having size \((k+1)\times(k+1)\). Applying the extended Cusick identity and using \eqref{nc:Phi_Cusick_relation}, we obtain:
\begin{align}
\det\mathcal{T}_{m+1}^{(\Phi)}
&=\pf\left[\mathcal{C}(u,v)-\mathcal{C}(q,p)\right]\nonumber\\
&=\pf\left[2M^{(m)}\begin{pmatrix}\omega^{(-s)}&-\boldsymbol{\sigma}\\\boldsymbol{\sigma}^{T}&0\end{pmatrix}\big(M^{(m)}\big)^{T}\right]\nonumber\\
&=2^{k+1}\det M^{(m)}\,\pf\begin{pmatrix}\omega^{(-s)}&-\boldsymbol{\sigma}\\\boldsymbol{\sigma}^{T}&0\end{pmatrix}.
\end{align}
The last matrix differs from \(\Omega^{(o)}\) only by the sign of its final row and column, so its Pfaffian is \(-\pf\big(\Omega^{(o)}\big)\). Since \(m\) is odd, \eqref{nc:Phi_M_det} also gives \(\det M^{(m)}=-(2m+1)\det\Lambda^{(m)}\). Hence, we obtain:
\begin{equation}
\det\mathcal{T}_{m+1}^{(\Phi)}=2^{k+1}(2m+1)\det\Lambda^{(m)}\,\pf\big(\Omega^{(o)}\big),\qquad m=2k+1.
\end{equation}
\end{proof}

\begin{proof}[Proof of Theorem \ref{intro:theorem_2}]
Using the relation \(\tau_{\ell}(\Psi)=(-p)^{\ell}\det\mathcal{T}_{\ell}^{(\Psi)}\), the two lemmas give:
\begin{equation}
\frac{\tau_{n}(\Phi)}{\tau_{n-1}(\Phi')}=
\begin{cases}
-p(2n-1)\dfrac{\pf\big(\Omega^{(e)}\big)}{\pf\big(\omega^{(s)}\big)}, & m=2k,\\
-2p(2n-1)\dfrac{\pf\big(\Omega^{(o)}\big)}{\pf\big(\widehat{\omega}^{(s)}\big)}, & m=2k+1.
\end{cases}
\end{equation}
Comparing these expressions with Theorem \ref{intro:theorem_1}, we obtain
\begin{equation}
\frac{\tau_{n}(\Phi)}{\tau_{n-1}(\Phi')}=-\frac{g_{n}}{f_{n}}=-\xi_{n},
\end{equation}
which proves Nakamura's conjecture.
\end{proof}

\section{Conclusion}

We have proved Nakamura's conjecture for the Yamazaki-Hori rational solutions of the Ernst equation for arbitrary \(n\). The proof establishes explicit identities between the Toda \(\tau\)-functions associated with the seeds \(\Phi\) and \(\Phi'\) and the Pfaffians representing the corresponding Ernst solutions. These identities hold for general values of \(p,q\) satisfying \(p^{2}+q^{2}=1\).

The main algebraic ingredient is an extension of the generalized Cusick identity to four arbitrary sequences. This identity expresses the Pfaffian of an antisymmetric matrix formed from convolution sums as the determinant of a block-Hankel matrix. We have also established a bordered version, needed for the odd-dimensional Toda matrices. Both identities hold independently of the Ernst equation and may be useful in establishing other determinant–Pfaffian correspondences.

An interesting question is whether the Pfaffian representations obtained in our previous work can be shown to satisfy the Ernst equation directly, without using their equivalence to the Yamazaki-Hori solutions. This would require deriving differential identities for the Pfaffians and using them to establish the bilinear equations associated with the Ernst equation.

Another question is whether other choices of the seed function \(\Phi\) in the 2D Toda molecule lead to solutions of the Ernst equation. It would be interesting to identify the conditions on \(\Phi\) under which such a correspondence exists.


\appendix
\section{Pfaffian forms for the polynomials \(f_{n}\) and \(g_{n}\)}
\label{app:appa}

In this appendix, we collect the explicit formulas for the antisymmetric matrices appearing in the Pfaffian representations of the Ernst polynomials \(f_n\) and \(g_n\), obtained in our previous work \cite{Melikyan2026pf}.\footnote{The corresponding formulas in \cite{Melikyan2026pf} contain additional terms proportional to \(\Delta=p^{2}+q^{2}-1\), which vanish on the physical subspace considered here.}

We begin with the matrices \(\omega\) and \(S\), defined by:
\begin{align}
    \omega_{ij}&:=\frac{1}{2}\int_{1}^{x} \left[\Gamma_{i}(w) \Gamma_{j}^{\prime}(w) -\Gamma_{i}^{\prime}(w) \Gamma_{j}(w) \right] \, dw
    +\frac{s^{2}}{2}\int_{1}^{y} \left[\Gamma_{i}(w) \Gamma_{j}^{\prime}(w) -\Gamma_{i}^{\prime}(w) \Gamma_{j}(w) \right] \, dw ,
    \label{mm:M_omega_part}\\
    S_{ij}&:=\frac{1}{2}\Gamma_{i}(x)\Gamma_{j}(x) + \frac{s^{2}}{2}\Gamma_{i}(y)\Gamma_{j}(y).
    \label{mm:M_S_part}
\end{align}
Here the polynomials \(\Gamma_i\) are given by:
\begin{align}
    \Gamma_{i}(w):=\frac{1}{2i-1}\left[w^{2i-1}-w^{2i+1}\right].
\end{align}
We further define the antisymmetric matrix \(\omega^{(s)}\):
\begin{align}
    \omega^{(s)}_{ij}:=\omega_{ij}-\mi s\,\widehat{\Gamma}_{ij},
    \qquad
    \widehat{\Gamma}_{ij}:=\frac{1}{2}\left[\Gamma_{i}(x)\Gamma_{j}(y)-\Gamma_{i}(y)\Gamma_{j}(x)\right].
\end{align}
For odd \(m=n-1\), we define the bordered antisymmetric matrix:
\begin{equation}
\widehat{\omega}^{(s)} =
\begin{pmatrix}
    \omega_{11}  & \cdots & \omega_{1m} & \Gamma_{1}^{(+)} \\
    \omega_{21}  & \cdots & \omega_{2m} & \Gamma_{2}^{(+)} \\
    \vdots & \vdots & \ddots & \vdots \\
    \omega_{m1}  & \cdots & \omega_{mm} & \Gamma_{m}^{(+)} \\[4pt]
    -\Gamma_{1}^{(+)} & \cdots & -\Gamma_{m}^{(+)} & 0
\end{pmatrix},
\label{pfodd:omega_s}
\end{equation}
where
\begin{align}
    \Gamma_{j}^{(+)}(x,y):=\Gamma_{j}(x)+\mi s\,\Gamma_{j}(y).
\end{align}
For even \(m=n-1\), the corresponding \((m+2)\times(m+2)\) antisymmetric matrix entering the Pfaffian representation of \(g_n\) is:
\begin{align}
    \Omega^{(e)}
    =
    \left(
    \begin{array}{cccc|cc}
    0 & \omega^{(s)}_{12} & \cdots & \omega^{(s)}_{1m} & \rho_1 & \beta_1 \\
    -\omega^{(s)}_{12} & 0 & \cdots & \omega^{(s)}_{2m} & \rho_2 & \beta_2 \\
    \vdots & \vdots & \ddots & \vdots & \vdots & \vdots \\
    -\omega^{(s)}_{1m} & -\omega^{(s)}_{2m} & \cdots & 0 & \rho_m & \beta_m \\[4pt]
    \hline
    -\rho_1 & -\rho_2 & \cdots & -\rho_m & 0 & \gamma \\
    -\beta_1 & -\beta_2 & \cdots & -\beta_m & -\gamma & 0
    \end{array}
    \right).
\end{align}

For odd \(m=n-1\), the corresponding antisymmetric matrix has the form:
\begin{align}
\Omega^{(o)}
=
\left(
\begin{array}{cccc|c}
0 & \omega^{(-s)}_{12} & \cdots & \omega^{(-s)}_{1m} & \sigma_1 \\
-\omega^{(-s)}_{12} & 0 & \cdots & \omega^{(-s)}_{2m} & \sigma_2 \\
\vdots & \vdots & \ddots & \vdots & \vdots \\
-\omega^{(-s)}_{1m} & -\omega^{(-s)}_{2m} & \cdots & 0 & \sigma_m \\[4pt]
\hline
-\sigma_1 & -\sigma_2 & \cdots & -\sigma_m & 0
\end{array}
\right).
\end{align}
Here the quantities \(\beta_{j}\), \(\rho_{j}\), \(\gamma\), and \(\sigma_{j}\) are defined by
\begin{align}
\beta_{j}&=\Gamma_{j}(x)-\mi s\Gamma_{j}(y),\qquad \gamma=\frac{x^{2m+1}-\mi s y^{2m+1}}{2m+1},\\
\rho_{j}&=\int_{1}^{x}\Gamma_{j}(w)w^{2m}\,dw+s^{2}\int_{1}^{y}\Gamma_{j}(w)w^{2m}\,dw,\\
\sigma_{j}&=\frac{\gamma}{2}\overline{\beta_{j}}-\rho_{j},\qquad 1\leq j\leq m.
\end{align}

\section{Proofs of the extended Cusick identities}
\label{app:appb}

\subsection{Proof of Theorem \ref{naka:ext_Cusick}}
We prove Theorem \ref{naka:ext_Cusick} by first establishing a convenient two-vector identity. Let
\begin{equation}
\mathbf{f}_{r}=\begin{pmatrix}
f^{(1)}_{r}\\
f^{(2)}_{r}
\end{pmatrix},\qquad \mathbf{g}_{r}=\begin{pmatrix}
g^{(1)}_{r}\\
g^{(2)}_{r}
\end{pmatrix},\qquad r\geq 1,
\end{equation}
be arbitrary sequences of two-dimensional vectors, and let \(\mathbf{A}=\mathbf{A}(\mathbf{f},\mathbf{g})\) be the \(2n\times 2n\) antisymmetric matrix with entries
\begin{equation}
[\mathbf{A}(\mathbf{f},\mathbf{g})]_{ij}=
\begin{cases}
\displaystyle\sum_{s=1}^{j-i}\mathbf{f}_{s+i-1}\cdot\mathbf{g}_{j-s}, & i<j,\\[1ex]
0, & i=j,\\
-[\mathbf{A}(\mathbf{f},\mathbf{g})]_{ji}, & i>j,
\end{cases}
\end{equation}
where
\begin{equation}
\mathbf{f}_{i}\cdot\mathbf{g}_{j} =f^{(1)}_{i} g^{(1)}_{j}+f^{(2)}_{i} g^{(2)}_{j}
\end{equation}
is the standard bilinear pairing. Also define
\begin{equation}
H^{(a)}_{f}=\bigl(f^{(a)}_{r+s-1}\bigr)_{r,s=1}^{n}, \qquad H^{(a)}_{g}=\bigl(g^{(a)}_{r+s-1}\bigr)_{r,s=1}^{n}, \qquad a=1,2.
\end{equation}
We claim that
\begin{equation}
\Phi(\mathbf{f},\mathbf{g}) :=\pf\mathbf{A}(\mathbf{f},\mathbf{g}) -\det\begin{pmatrix}
H^{(1)}_{f} & H^{(2)}_{g}\\
-H^{(2)}_{f} & H^{(1)}_{g}
\end{pmatrix}\equiv 0.
\label{VectorDale}
\end{equation}
We prove this below by polarization and atomic parametrization. For the algebraic arguments below, let \(\mathbb K\) be a field of characteristic zero. All sequences, vectors, and matrices in this appendix are understood to have entries in \(\mathbb K\). The application to the Ernst problem is obtained by taking \(\mathbb K=\mathbb C\).

\begin{proof}[Proof of the auxiliary vector identity]
The proof is based on two ingredients.

First, the matrix \(\mathbf{A}(\mathbf{f},\mathbf{g})\) is bilinear in \(\mathbf{f}\) and \(\mathbf{g}\). Consequently, \(\pf\mathbf{A}(\mathbf{f},\mathbf{g})\) is bihomogeneous of degree \((n,n)\) in \((\mathbf{f},\mathbf{g})\). The determinant in the second term of \eqref{VectorDale} has the same bidegree: its first \(n\) columns are linear in \(\mathbf{f}\), while its last \(n\) columns are linear in \(\mathbf{g}\). Thus \(\Phi\) is bihomogeneous of bidegree \((n,n)\).

We define its full polarization by
\begin{equation}
\widetilde{\Phi}\bigl(\mathbf{F}_{(1)},\ldots,\mathbf{F}_{(n)};\mathbf{G}_{(1)},\ldots,\mathbf{G}_{(n)}\bigr)
:= [t_{1}\cdots t_{n}s_{1}\cdots s_{n}]\,\Phi\left(\sum_{\alpha=1}^{n}t_{\alpha}\mathbf{F}_{(\alpha)},\sum_{\beta=1}^{n}s_{\beta}\mathbf{G}_{(\beta)}\right),
\end{equation}
where \([t_1\cdots t_n s_1\cdots s_n]\) denotes the coefficient of the product of all the distinct parameters. The polarized form is multilinear in all of its \(2n\) arguments, and
\begin{equation}
\widetilde{\Phi}(\mathbf{f},\ldots,\mathbf{f}; \mathbf{g},\ldots,\mathbf{g}) =(n!)^2\Phi(\mathbf{f},\mathbf{g}).
\end{equation}
Consequently, it is enough to prove that \(\widetilde{\Phi}\equiv0\).

The second ingredient is the moment atomic representation. Put \(m=2n-1\), and choose arbitrary distinct nodes \(z_1,\ldots,z_m\). The Vandermonde matrix
\begin{equation}
V(z)=\bigl(z_{r}^{k-1}\bigr)_{k,r=1}^{m}
\end{equation}
is invertible. Therefore every vector-valued moment sequence \(\{\mathbf{f}_k\}_{k=1}^{m}\) has an atomic representation
\begin{equation}
\mathbf{f}_{k}=\sum_{r=1}^{m}\mathbf{p}_{r} z_{r}^{k-1},
\end{equation}
with uniquely determined vector weights \(\mathbf{p}_r\), for the chosen nodes. Hence the vector atoms
\begin{equation}
\mathfrak{a}(x,\mathbf{u}) :=\bigl(\mathbf{u},\mathbf{u}x,\ldots, \mathbf{u}x^{2n-2}\bigr), \qquad \mathbf{u}\in \mathbb{K}^2,
\end{equation}
span the space of all moment sequences. Since \(\widetilde{\Phi}\) is multilinear, it is therefore enough to evaluate it when each of its arguments is one atom.

Choose atoms \(\mathfrak a(x_\alpha,\mathbf u_\alpha),\
\mathfrak a(y_\beta,\mathbf v_\beta),
\
\alpha,\beta=1,\ldots,n,
\)
and define
\begin{equation}
\mathbf{F}_{k}(\mathbf{t}) :=\sum_{\alpha=1}^{n} t_{\alpha}\mathbf{u}_{\alpha} x_{\alpha}^{k-1}, \qquad \mathbf{G}_{k}(\mathbf{s}) :=\sum_{\beta=1}^{n} s_{\beta}\mathbf{v}_{\beta} y_{\beta}^{k-1}.
\end{equation}
It is enough to prove that \(\Phi(\mathbf{F}(\mathbf{t}),\mathbf{G}(\mathbf{s}))\) vanishes identically in the parameters \(t_\alpha,s_\beta\). The coefficient of \(t_1\cdots t_n s_1\cdots s_n\) will then give the desired polarized identity.

For \(i<j\), we have
\begin{align}
[\mathbf{A}(\mathbf{F}(\mathbf{t}),\mathbf{G}(\mathbf{s}))]_{ij} &= \sum_{\alpha,\beta=1}^{n}t_{\alpha}s_{\beta}(\mathbf{u}_{\alpha}\cdot\mathbf{v}_{\beta})\sum_{r=1}^{j-i}x_{\alpha}^{r+i-2}y_{\beta}^{j-r-1}\nonumber\\
&= \sum_{\alpha,\beta=1}^{n}t_{\alpha}s_{\beta}(\mathbf{u}_{\alpha}\cdot\mathbf{v}_{\beta})\frac{x_{\alpha}^{i-1}y_{\beta}^{j-1}-x_{\alpha}^{j-1}y_{\beta}^{i-1}}{y_{\beta}-x_{\alpha}},
\label{A}
\end{align}
where in the last equality we used the sum of a geometric progression.

Since the atom weights are arbitrary, we may absorb the parameters \(t_\alpha\) and \(s_\beta\) into them for the generic calculation below. After the resulting identity has been established for arbitrary weights, we restore the original parameters and extract the coefficient of \(t_1\cdots t_n s_1\cdots s_n\). Thus, from now on, write
\begin{equation}
t_{\alpha}\mathbf{u}_{\alpha}=\begin{pmatrix}
u^{(1)}_{\alpha}\\
u^{(2)}_{\alpha}
\end{pmatrix}, \qquad s_{\beta}\mathbf{v}_{\beta}=\begin{pmatrix}
v^{(1)}_{\beta}\\
v^{(2)}_{\beta}
\end{pmatrix},
\end{equation}
and relabel these absorbed weights again as \(\mathbf{u}_\alpha\) and \(\mathbf{v}_\beta\). Define the diagonal matrices
\begin{equation}
P_{a}:=\operatorname{diag}\bigl(u^{(a)}_{1},\ldots,u^{(a)}_{n}\bigr), \qquad Q_{a}:=\operatorname{diag}\bigl(v^{(a)}_{1},\ldots,v^{(a)}_{n}\bigr), \qquad a=1,2,
\end{equation}
as well as the Cauchy matrix
\begin{equation}
C_{\alpha\beta}:=\frac{1}{y_{\beta}-x_{\alpha}}.
\end{equation}
Then
\begin{equation}
\Gamma_{\alpha\beta} :=\frac{\mathbf{u}_{\alpha}\cdot\mathbf{v}_{\beta}} {y_{\beta}-x_{\alpha}}
\end{equation}
can be written as
\begin{equation}
\Gamma=P_{1}CQ_{1}+P_{2}CQ_{2}.
\end{equation}

Introduce the \(n\times n\) Vandermonde matrices
\begin{equation}
(X)_{r\alpha}:=x_{\alpha}^{r-1}, \qquad (Y)_{r\beta}:=y_{\beta}^{r-1}, \qquad r=1,\ldots,n,
\end{equation}
and the \(2n\times n\) extended Vandermonde matrices
\begin{equation}
(\widehat X)_{i\alpha}:=x_{\alpha}^{i-1}, \qquad (\widehat Y)_{i\beta}:=y_{\beta}^{i-1}, \qquad i=1,\ldots,2n.
\end{equation}
Finally, put
\begin{equation}
\mathcal W:=(\widehat X\;\widehat Y).
\end{equation}
Using \eqref{A}, we obtain
\begin{equation}
\mathbf{A}(\mathbf{F},\mathbf{G}) =\mathcal W \mathcal K\mathcal W^{T},\qquad \mathrm{where}\qquad \mathcal K:=\begin{pmatrix}
0 & \Gamma\\
-\Gamma^{T} & 0
\end{pmatrix}.
\end{equation}
Therefore, using the standard Pfaffian congruence property,
\begin{equation}
\pf\mathbf{A} =\det\mathcal W\,\pf \mathcal K =(-1)^{\frac{n(n-1)}2} \det\mathcal W\,\det\Gamma.
\end{equation}

Let
\begin{equation}
\Delta(x):=\prod_{\alpha<\alpha'}(x_{\alpha'}-x_{\alpha}), \qquad \Delta(y):=\prod_{\beta<\beta'}(y_{\beta'}-y_{\beta}).
\end{equation}
Since all the \(x\)-columns precede all the \(y\)-columns, we get
\begin{equation}
\det\mathcal W =\Delta(x)\Delta(y) \prod_{\alpha,\beta=1}^{n}(y_{\beta}-x_{\alpha}).
\end{equation}
The Cauchy determinant formula, with our convention for \(C\), is
\begin{equation}
\det C =(-1)^{\frac{n(n-1)}2} \frac{\Delta(x)\Delta(y)} {\displaystyle\prod_{\alpha,\beta=1}^{n}(y_{\beta}-x_{\alpha})}.
\end{equation}

For the moment, assume that \(P_1,Q_1,X,Y\), and \(C\) are invertible. Put
\begin{equation}
D:=P_{1}^{-1}P_{2}, \qquad E:=Q_{2}Q_{1}^{-1}.
\end{equation}
Then
\begin{equation}
\det\Gamma =\det\bigl( P_{1}C\bigl(\mathds{1}+C^{-1}DCE\bigr)Q_{1}\bigr)=\det P_{1}\det Q_{1}\det C\, \det\bigl(\mathds{1}+C^{-1}DCE\bigr).
\end{equation}
Combining the last three equations gives
\begin{equation}
\pf\mathbf{A} =\Delta(x)^2\Delta(y)^2 \det P_{1}\det Q_{1}\, \det\bigl(\mathds{1}+C^{-1}DCE\bigr).
\label{PfAtomic}
\end{equation}

We now calculate the determinant in the second term of \eqref{VectorDale}. By construction,
\begin{equation}
H^{(a)}_{f}=XP_{a}X^{T}, \qquad H^{(a)}_{g}=YQ_{a}Y^{T}.
\end{equation}
Therefore
\begin{equation}
\det\begin{pmatrix}
H^{(1)}_{f} & H^{(2)}_{g}\\
-H^{(2)}_{f} & H^{(1)}_{g}
\end{pmatrix}= \det(X)^2\det(Y)^2\det R = \Delta(x)^2\Delta(y)^2 \det R,
\label{HHHH}
\end{equation}
where
\begin{equation}
R:=\begin{pmatrix}
P_{1} & X^{-1}YQ_{2}\\
-Y^{-1}XP_{2} & Q_{1}
\end{pmatrix}.
\end{equation}

It remains to compare \(\det R\) with the last factor in \eqref{PfAtomic}. Set
\begin{equation}
S:=Y^{-1}X, \qquad \Lambda:=\operatorname{diag}(\lambda_{1},\ldots,\lambda_{n}), \qquad \lambda_{\beta}:=\prod_{\alpha=1}^{n}(y_{\beta}-x_{\alpha}),
\end{equation}
where \(S_{\beta\gamma}\) are the Lagrange interpolation weights \(\ell_\beta(z)\), evaluated at \(z = x_\gamma\)
\begin{equation}
\ell_{\beta}(z):= \prod_{\delta\neq\beta} \frac{z-y_{\delta}}{y_{\beta}-y_{\delta}},\ \ S_{\beta\gamma} =\prod_{\delta\neq\beta} \frac{x_{\gamma}-y_{\delta}}{y_{\beta}-y_{\delta}}\equiv \ell_{\beta}(x_{\gamma}).
\end{equation}
By the Lagrange interpolation formula
\begin{equation}
p_{\alpha}(z):=\prod_{\rho\neq\alpha}(z-x_{\rho}) = \sum_{\beta=1}^n p_{\alpha}(y_{\beta})\,\ell_{\beta}(z) \equiv \sum_{\beta=1}^n\prod_{\rho\neq\alpha}(y_{\beta}-x_{\rho})\,\ell_{\beta}(z).
\end{equation}
By evaluating this at \(z = x_\gamma\), we get the following elementary interpolation identity:
\begin{equation}
(C\Lambda S)_{\alpha\gamma} = \sum_{\beta=1}^{n} \frac{\lambda_{\beta}}{y_{\beta}-x_{\alpha}} S_{\beta\gamma} = \sum_{\beta=1}^{n} p_{\alpha}(y_{\beta})\ell_{\beta}(x_{\gamma}) = p_{\alpha}(x_{\gamma})
\end{equation}
or
\begin{equation}
C\Lambda S =\operatorname{diag}\left( \prod_{\rho\neq1}(x_{1}-x_{\rho}),\ldots, \prod_{\rho\neq n}(x_{n}-x_{\rho}) \right).
\label{Interp}
\end{equation}
Since the diagonal matrix on the right-hand side of \eqref{Interp} commutes with \(D\), we obtain
\begin{equation}
SDS^{-1}=\Lambda^{-1}C^{-1}DC\Lambda.
\label{Conjugacy}
\end{equation}
Using the Schur complement of \(P_1\), we find
\begin{equation}
\det R =\det P_{1}\det\bigl(Q_{1}+SDS^{-1}Q_{2}\bigr)= \det P_{1}\det Q_{1} \det\bigl(\mathds{1}+SDS^{-1}E\bigr).
\end{equation}
By \eqref{Conjugacy}, and since both \(E\) and \(\Lambda\) are diagonal,
\begin{equation}
\det\bigl(\mathds{1}+SDS^{-1}E\bigr) =\det\Bigl( \Lambda^{-1}\bigl(\mathds{1}+C^{-1}DCE\bigr)\Lambda \Bigr) =\det\bigl(\mathds{1}+C^{-1}DCE\bigr).
\end{equation}
Therefore
\begin{equation}
\det R =\det P_{1}\det Q_{1} \det\bigl(\mathds{1}+C^{-1}DCE\bigr).
\end{equation}
Thus \eqref{HHHH} is equal to \eqref{PfAtomic}. Hence
\begin{equation}
\Phi(\mathbf{F}(\mathbf{t}),\mathbf{G}(\mathbf{s}))=0
\end{equation}
for arbitrary atomic data.

The identity was derived for generic choices for which the intermediate matrices are invertible. Since both sides are polynomial in the underlying parameters, and \(\mathbb K\) is infinite, the identity extends to all choices. Finally, extracting the coefficient of \(t_1 t_2\cdots t_n s_1 s_2\cdots s_n\) gives \(\widetilde{\Phi}=0\) on all atomic arguments. Since the atoms span the moment space and \(\widetilde{\Phi}\) is multilinear, it vanishes identically. The diagonal relation for the polarization then gives \(\Phi(\mathbf{f},\mathbf{g})=0\) for arbitrary sequences.
\end{proof}

To recover Theorem \ref{naka:ext_Cusick}, set \(n=k\) and
\begin{equation}
\mathbf{f}_{j}=\begin{pmatrix}
u_{j-1}\\
-q_{j-1}
\end{pmatrix}, \qquad \mathbf{g}_{j}=\begin{pmatrix}
v_{j-1}\\
p_{j-1}
\end{pmatrix}, \qquad j=1,\ldots,2k-1.
\end{equation}
Then
\begin{equation}
\mathbf{A}(\mathbf{f},\mathbf{g}) =\mathcal{C}(u,v)-\mathcal{C}(q,p),
\end{equation}
while the block determinant in \eqref{VectorDale} is
\begin{equation}
\det\begin{pmatrix}
\mathcal{U} & \mathcal{P}\\
\mathcal{Q} & \mathcal{V}
\end{pmatrix}.
\end{equation}
This proves Theorem \ref{naka:ext_Cusick}. Finally, setting \(p_j=q_j=0\) and using \(u_j=\phi_{j+1}\), \(v_j=\psi_{j+1}\) gives Theorem \ref{nc:Dale_Cusick_theorem}, the generalized Cusick identity.

\subsection{Proof of Theorem \ref{naka:bordered_Cusick}}

Set
\begin{equation}
n=k+1,\qquad m=2k+1=2n-1.
\end{equation}
Extend \(p,q,v\) by zero:
\begin{equation}
\widetilde p_{2k}=\widetilde q_{2k} =\widetilde v_{2k}=0,
\end{equation}
while \(\widetilde p_j=p_j\), etc., for \(0\le j\le 2k-1\).

Introduce an auxiliary parameter \(t\) and define the two-dimensional sequences required for the even theorem by
\begin{equation}
\mathbf f_{r}=\begin{pmatrix}
u_{r-1}\\
-\widetilde q_{r-1}
\end{pmatrix}, \qquad \mathbf g_{r}(t)=\begin{pmatrix}
\widetilde v_{r-1}+t\,\delta_{r,\,2k+1}\\
\widetilde p_{r-1}
\end{pmatrix}, \qquad r=1,\ldots,2k+1.
\end{equation}
The signs are chosen precisely so that
\begin{equation}
\mathbf f_{r}\cdot\mathbf g_{s}(t) = u_{r-1}\widetilde v_{s-1} -\widetilde q_{r-1}\widetilde p_{s-1} +t\,u_{r-1}\delta_{s,\,2k+1}.
\end{equation}
Let \(\mathbf A(t)\) be the matrix in the auxiliary vector identity \eqref{VectorDale}, with \(n=k+1\), constructed from these sequences. The upper-left \(m\times m\) block is exactly
\begin{equation}
C_{m}=\mathcal C(u,v)-\mathcal C(q,p).
\end{equation}
Moreover, the parameter \(t\) occurs only through the last moment of the first component of \(\mathbf g\). Therefore it can only affect the last column and row.
\begin{equation}
A_{i+1,\,2k+2}(t)=c_{i}+t\,u_{i},
\end{equation}
where
\begin{equation}
c_{i} = \sum_{s=1}^{2k-i} \left( u_{i+s}\,\widetilde v_{\,2k-s} - \widetilde q_{\,i+s}\,\widetilde p_{\,2k-s} \right), \qquad i=0,\ldots,2k,
\end{equation}
where the sum is empty for \(i=2k\), so that \(c_{2k}=0\). Thus
\begin{equation}
\mathbf A(t)=\begin{pmatrix}
C_{m} & \mathbf c+t\mathbf u\\
-(\mathbf c+t\mathbf u)^{T} & 0
\end{pmatrix}.
\end{equation}
Since the Pfaffian is linear in the bordering vector,
\begin{equation}
[t]\,\pf\mathbf A(t) = \pf\begin{pmatrix}
C_{m} & \mathbf u\\
-\mathbf u^{T} & 0
\end{pmatrix}.
\label{appb_1}
\end{equation}
Applying the auxiliary vector identity \eqref{VectorDale} with \(n=k+1\), we obtain
\begin{equation}
\pf \mathbf A(t)=\det\begin{pmatrix}
\widehat H(u) & \widehat H(\widetilde p)\\
\widehat H(\widetilde q) & \widehat H(\widetilde v)+tE
\end{pmatrix},
\end{equation}
where, for a sequence \(a\),
\begin{equation}
\widehat H(a):=\bigl(a_{i+j}\bigr)_{i,j=0}^{k}
\end{equation}
is the \((k+1)\times(k+1)\) Hankel matrix, and \(E\) is the \((k+1)\times(k+1)\) matrix with only its lower-right entry equal to \(1\). Indeed, the four blocks arise as
\begin{equation}
H_{f}^{(1)}=\widehat H(u),\qquad H_{g}^{(2)}=\widehat H(\widetilde p),\qquad -H_{f}^{(2)}=\widehat H(\widetilde q),\qquad H_{g}^{(1)}=\widehat H(\widetilde v)+tE,
\end{equation}
where the sign in the third block follows from \(\mathbf f^{(2)}=-\widetilde q\). The parameter \(t\) appears only in the lower-right corner of the full determinant. Deleting the last row and column leaves \(\mathcal U\), the first \(k\) columns of \(\widehat H(\widetilde p)\), the first \(k\) rows of \(\widehat H(\widetilde q)\), and the upper-left \(k\times k\) block of \(\widehat H(\widetilde v)\), namely \(\mathcal P,\mathcal Q,\mathcal V\), respectively. Hence:
\begin{equation}
[t]\det\begin{pmatrix}
\widehat H(u) & \widehat H(\widetilde p)\\
\widehat H(\widetilde q) & \widehat H(\widetilde v)+tE
\end{pmatrix}= \det\begin{pmatrix}
\mathcal U & \mathcal P\\
\mathcal Q & \mathcal V
\end{pmatrix}.
\label{appb_2}
\end{equation}
Taking the coefficient of \(t\) in the even identity and using \eqref{appb_1} and \eqref{appb_2}, we obtain
\begin{equation}
\pf\begin{pmatrix}
C_{m} & \mathbf u\\
-\mathbf u^{T} & 0
\end{pmatrix}= \det\begin{pmatrix}
\mathcal U & \mathcal P\\
\mathcal Q & \mathcal V
\end{pmatrix}.
\end{equation}
This completes the proof of Theorem \ref{naka:bordered_Cusick}.

\bibliographystyle{elsarticle-num}
\bibliography{ncfinal}
\end{document}